LLNL-PROC-643794

LAWRENCE LIVERMORE NATIONAL LABORATORY

# X-ray Thomson scattering for partially ionized plasmas including the effect of bound levels

J. Nilsen, W. R. Johnson, K. T. Cheng

**Disclaimer**

# X-ray Thomson scattering for partially ionized plasmas including the effect of bound levels

Joseph Nilsen[1], Walter R. Johnson[2], and K. T. Cheng[1]

[1]Lawrence Livermore National Laboratory, Livermore, CA 94551
[2]University of Notre Dame, Notre Dame, IN 46556

## ABSTRACT

X-ray Thomson scattering is being developed as a method to measure the temperature, electron density, and ionization state of high energy density plasmas such as those used in inertial confinement fusion. Most experiments are currently done at large laser facilities that can create bright X-ray sources, however the advent of the X-ray free electron laser (X-FEL) provides a new bright source to use in these experiments. One challenge with X-ray Thomson scattering experiments is understanding how to model the scattering for partially ionized plasmas in order to include the contributions of the bound electrons in the scattered intensity. In this work we take the existing models of Thomson scattering that include elastic ion-ion scattering and the electron-electron plasmon scattering and add the contribution of the bound electrons in the partially ionized plasmas. We validated our model by analyzing existing beryllium experimental data. We then consider several higher Z materials such as Cr and predict the existence of additional peaks in the scattering spectrum that requires new computational tools to understand. We also show examples of experiments in CH and Al that have bound contributions that change the shape of the scattered spectra.

## 1. INTRODUCTION

For warm dense plasmas, X-ray Thomson scattering is being used as an important diagnostic technique to measure temperatures, densities, and ionization balance. Glenzer and Redmer [1] have reviewed the underlying theory of Thomson scattering used to analyze experiments.

In this paper we start with the theoretical model proposed by Gregori et al. [2] but evaluate the Thomson-scattering dynamic structure function using parameters taken from our own average-atom code [3,4]. The average-atom model is a quantum mechanical version of the temperature-dependent Thomas-Fermi model of plasma developed years ago by Feynman et al. [5]. It consists of a single ion of charge Z with a total of Z bound and continuum electrons in a Wigner-Seitz cell that is embedded in a uniform "jellium sea" of free electrons whose charge $Z_f$ is balanced by a uniform positive background. This model enables one to consider the contributions from the bound electrons in a self-consistent way for any ion. Approaches such as Gregori's employ hydrogenic wave functions with screening factors to approximate the contribution from the bound electrons for a limited number of materials and requires the user to specify the ionization balance.

To validate our average-atom code we analyzed existing experimental data [6] for beryllium. This is discussed in Ref. [7]. For that Be case the bound electrons were not important. In this paper we consider several higher Z materials, such as Cr, CH, and Al, and predict the existence of additional peaks in the scattering spectrum that requires new computational tools to understand.

## 2. THEORY OVERVIEW

Refs. [3-4] provide a more detailed description of the theory used in this paper. To summarize, the Thomson scattering cross section is proportional to the dynamic structure function $S(\hbar k, \hbar\omega)$, where $\hbar k$ and $\hbar\omega$ are the momentum and energy transfers, respectively, from the incident to the scattered photons. For simplicity, $\hbar$ is dropped in the formuli. The work of Chihara [8,9] showed that $S(k,\omega)$ can be decomposed into three terms: the first term $S_{ii}(k,\omega)$ is the contribution from elastic scattering by electrons that follow the ion motion, the second term $S_{ee}(k,\omega)$ is the contribution from inelastic scattering by free electrons, and the third term $S_b(k,\omega)$ is the inelastic scattering contribution by bound electrons from bound-free transitions modulated by the ionic motion. In the present work, the modulation factor is ignored when evaluating the bound-free scattering structure function. For the bound-free contribution, our calculations use average-atom scattering wave functions for the final states. In Ref. [4] we showed that using plane-wave final states can give very different results that disagree with experimental data.

The theoretical model developed by Gregori et al. [2] is used to evaluate the ion-ion contribution $S_{ii}(k,\omega)$ to the dynamic structure function, with scattering form factors calculated with bound- and continuum-state wave functions from our average-atom code. The procedure proposed in Ref. [10] is used to account for differences between electron and ion temperatures. The electron-electron contribution $S_{ee}(k,\omega)$ is expressed in terms of the dielectric function $\varepsilon(k,\omega)$ of the free electrons, which in turn is evaluated using the random-phase approximation (RPA) as in Ref. [2]. Finally, bound-state contributions to the dynamic structure function are evaluated using average-atom bound- and continuum-state wave functions, with the latter approaching plane waves asymptotically.

In the collective regime, which is for small momentum transfers and usually for forward angle scattering, one sees plasmon peaks that are up and down shifted in energy from the central peak. In experiments that can observe both peaks, the plasma temperature is determined from the ratio of the two plasmon peaks by $\exp[-\Delta E/kT]$ where $\Delta E$ is the energy shift of the plasmon peak [1,4]. The energy shift of the plasmon peak has a plasma frequency component that is proportional to the square root of the free-electron density but it also includes a thermal energy contribution and a Compton shift that depend on the x-ray energy, temperature, and scattering angle [1]. As a result, the electron density is usually determined by doing a best fit to the experimental data.

## 3. MODELLING EXPERIMENTS

Now lets consider a higher-Z materials such as Cr at an electron temperature of 10 eV where the bound state contribution can be very important. The average-atom code predicts that Cr at the solid density of 7.19 g/cc heated to 10 eV has 6.2 continuum electrons in the Wigner-Seitz cell. This is a nearly closed Ar-like core consisting of almost fully occupied 3s and 3p subshells with binding energies of 57.7 and 30.7 eV, respectively. For the asymptotic value of the free electrons as one goes to large radius the average-atom code predicts $Z_f = 2.92$ and an electron density of $2.4 \times 10^{23}$ per cc. This is a case where there is a factor of 2 difference between the number of continuum electrons in the Wigner-Seitz cell and the asymptotic value of the free electrons, $Z_f$, in the jellium sea. For codes such as Gregori's the number of free electrons is an input parameter so it is important to understand what is the best value to choose. We think this is a very important issue to understand because we use the asymptotic value $Z_f$ to determine the free electrons that contribute to $S_{ee}$ (plasmon peak) and assume that some of the continuum electrons are bunched near the ion and follow the ionic motion and do not act as if they are free electrons. In the average-atom code, the electron density is the product of the ion density and $Z_f$.

Figure 1 shows the scattered intensity versus photon energy for calculations for Cr done with (solid line) and without (dashed line) the contribution of bound electrons for a 4750 eV X-ray source scattered at 40° off Cr. Without bound electrons, we predict the central elastic scattering peak from $S_{ii}(k,\omega)$ and the plasmon peaks from $S_{ee}(k,\omega)$. When we include the effect of bound electrons, we predict a very strong scattering peak that is downshifted by about 40 eV from the central elastic peak due to the 3p electrons and a much weaker 3s peak at lower energy.

Recent experiments were done at the Omega Laser Facility [11] in Rochester to look at Thomson scattering off compressed CH at an estimated density 8 g/cc with a 9.0 keV source from He-α Zn at a scattering angle of 135 degrees. The CH is compressed by three converging coalescing shocks generated by 8 beams of the Omega laser. The experiments estimate an electron temperature of 10 eV with an electron density of 1.4 x $10^{24}$ $cm^{-3}$ at a time of 3.4 ns. The assumption in the fitting calculation using Gregori's code is that the carbon has 2 bound electrons from the K shell and the other electrons in carbon and hydrogen are ionized and counted as free electrons. To model this problem with the average-atom code we had to address the issue of mixtures. To do that we assumed CH had a certain density and then had to find the local density for C and H independently such that the electron density and chemical potential would be the same for both materials. To get an electron density of 1.4 x $10^{24}$ $cm^{-3}$ we needed a total density for CH of 8.896 g/cc. At this density the average-atom code calculates that $Z_f$ = 2.67 for C and 0.73 for H. However the average-atom code predicts that only the K shell of C is bound so there are 4 continuum electrons in C and 1 in H, which is the assumption used in the fitting calculations done in Ref. [11]. However if one assumes the continuum electrons are all free electrons than the matter density is 6 g/cc, which differs significantly from the value calculated by the average-atom code. Figure 2 shows the scattered intensity versus photon energy for the experiment [11] and our average-atom code calculations. The dots are the experimental data. The thick solid line is the average-atom calculation. Also shown are the three contributions to the total scattering. The elastic peak is fit to the experiment and shown by a thin solid line centered at 9 keV. The dashed line shows the free-free contribution centered at 8770 eV. The dotted line shows the bound-free contribution and it starts rising about 8850 eV. One can see that the bound-free contribution from the K-shell electrons dominates the low energy wing of the scattered line.

Other recent experiments were done at the Omega-60 laser facility [12] to look at Thomson scattering off compressed Al at an estimated density 8.1 g/cc with a 17.9. keV source from He-α Mo at a scattering angle of 69 degrees. The Al is compressed by a single strong shock generated by 9 beams of the Omega laser. Figure 3 shows the scattered intensity versus photon energy for the experiment [12] and calculations. The experiment is shown by the noisy dotted line. The thick solid line shows the fit to the Gregori code with the various contributions shown as dashed and dashed-dotted lines. For the average-atom code we only show the bound-free contribution as a dashed line since both codes agree for the free-free contribution and the ion-ion elastic contribution is just a fit to the data. The main point is that the bound-free contribution for the average-atom code falls off much more quickly than Gregori's code as one goes to lower energy. While the data is too noisy in the wings to distinguish between the two theories we believe our use of distorted wave functions for the free electrons causes the behavior to be much different than Gregori's code in the wings. In the future, experiments with a stronger source such as the XFEL should enable us to distinguish between the different theoretical approaches.

## 4. CONCLUSIONS

The availability of bright monochromatic X-ray line sources from X-ray free electron laser facilities opens up many new possibilities to use Thomson scattering as a diagnostic technique to measure the temperatures, densities, and ionization balance in warm dense plasmas.

Current attempts to model Thomson scattering tend to use simplified models to estimate the effect of the bound electrons on the measured scattered intensity. Our approach here is to evaluate the Thomson-scattering dynamic structure function using parameters taken from our average-atom code [3,4]. This model enables us to consider contributions from the bound electrons in a self-consistent way for any ion.

We validate our average-atom based Thomson scattering code by comparing our model against existing experimental data for Be, in conditions near solid density and temperature near 18 eV. For solid density Cr at 10 eV we predict the existence of additional peaks in the scattering spectrum, the understanding of which requires new computational tools. We also analyze compressed CH and Al at densities above solid density and temperatures of 10 eV. We show the importance of the bound-free contributions to the low energy wing of the scattered spectra. We also discuss that there are still significant differences in the theory that need to be understood. One of the big issues that needs to be understood is the difference between the non-bound continuum electrons and the free electrons and how these contribute to the different terms in the scattering.

## ACKNOWLEDGEMENTS

We would like to thank Luke Fletcher, Tammy Ma, Tilo Döppner, and Siegfried Glenzer for fruitful discussions and access to the experimental data. This work was performed under the auspices of the U.S. Department of Energy by Lawrence Livermore National Laboratory under Contract DE-AC52-07NA27344.

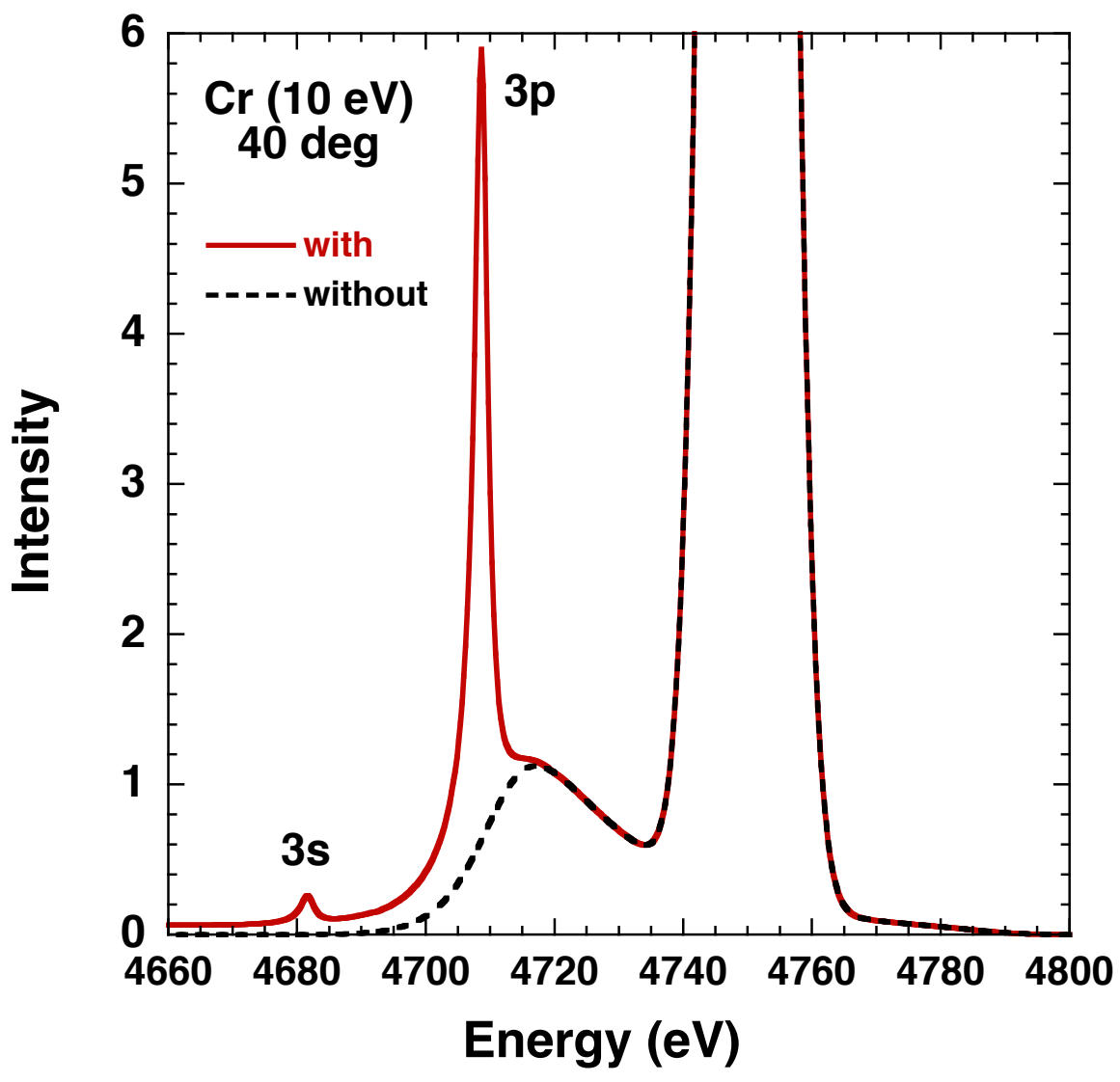

**Fig. 1.** Intensity vs scattered photon energy for calculation of scattering of a 4750 eV X-ray line from Cr at 40° for an electron temperature of 10 eV. The case where the 3p and 3s bound electron contributions are included is shown by the solid line and the case without the bound electrons is shown by the dashed line.

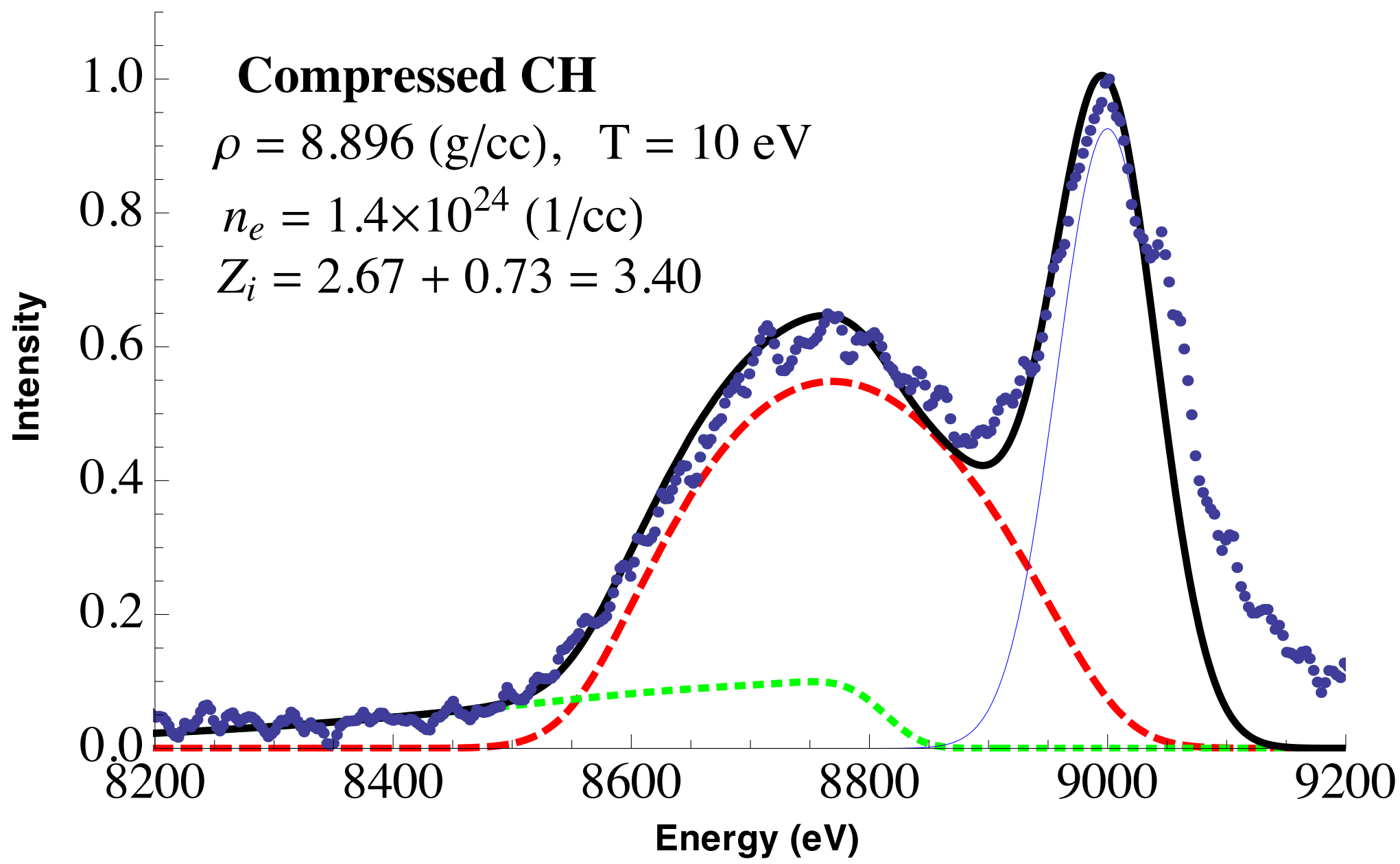

**Fig. 2.** Intensity vs scattered photon energy for scattering of a 9 keV X-ray line from compressed CH at 135° for an electron temperature of 10 eV. The solid thick line is the average-atom code calculation. Also shown are the three components that contribute to the scattering term. The solid thin line is the fit to the elastic peak. The dashed line (red-online) shows the free-free contribution centered at 8770 eV. The dotted line (green-online) shows the bound-free contribution from the K-shell start to rise about 8850 eV and dominate the low energy wing of the scattered line.

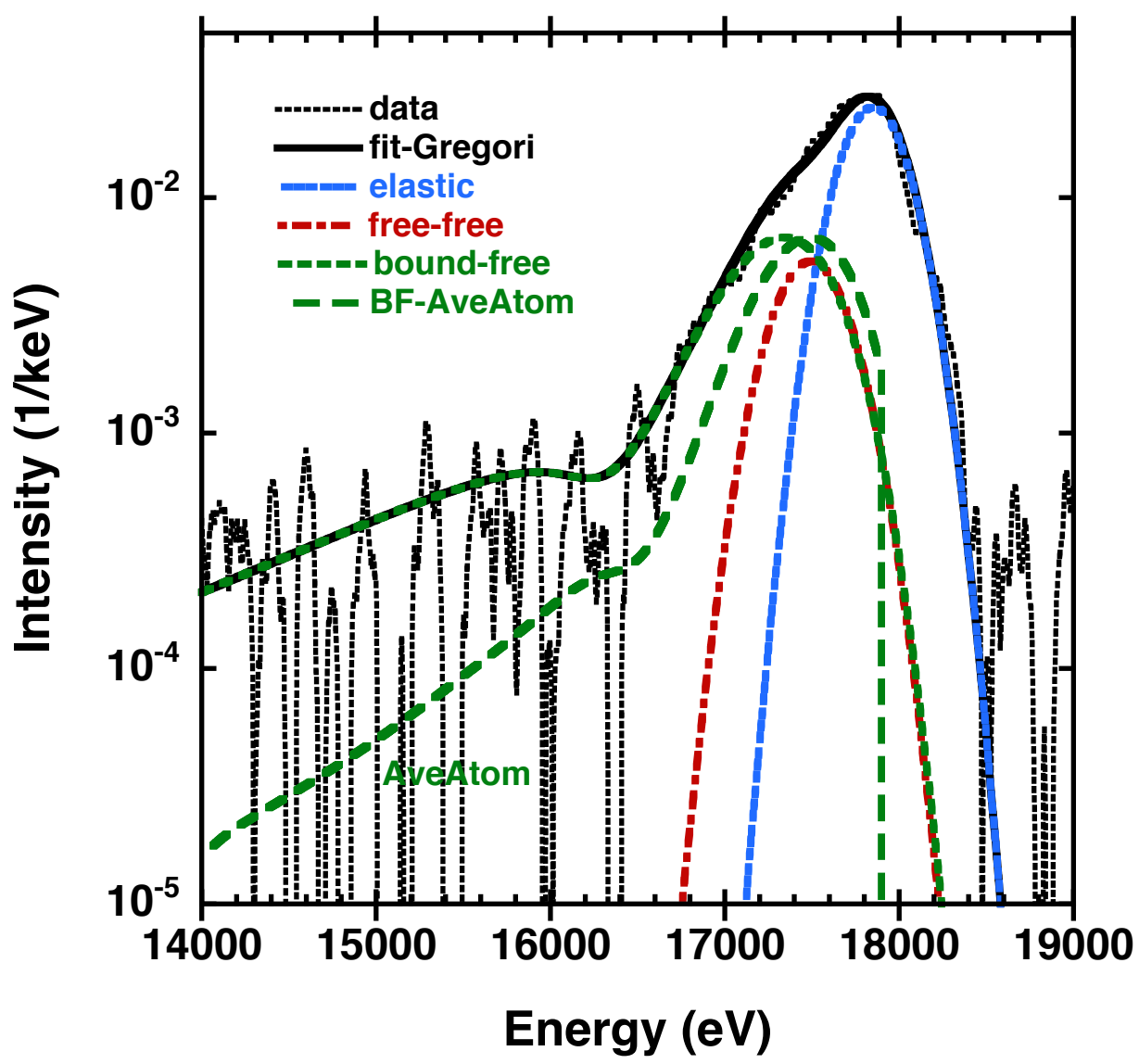

**Fig. 3.** Intensity vs scattered photon energy scattering of a 17.9 keV X-ray line from compressed Al at 69° for an electron temperature of 10 eV. The solid thick line is the Gregori code calculation along with the 3 components shown as dashed and dashed-dotted lines. The bound-free contribution from the average-atom code is shown as a thick dashed line (green-online) that falls off much more rapidly in the low energy wing of the line than the Gregori code result.